# The European Union Deforestation Regulation: The Impact on Argentina

Pablo de la Vega[1]

This version: August, 2025

**Abstract**

We analyze the potential economic impacts in Argentina of the European Union Deforestation Regulation (EUDR), which as of January 2026 will prohibit the export to the European Union of certain raw materials and related products if they involve the use of deforested land. We estimate that the EUDR would cover around 6 billion US dollars in exported value, but only 2.84% is not compliant with the EUDR, with soy and cattle being the most affected production chains. We use a dynamic computable general equilibrium model to simulate the impact of the EUDR on the Argentine economy. If the non-compliant production cannot enter the EU market because of the EUDR, the results of the simulations suggest that the potential macroeconomic impacts are limited: GDP would be reduced by an average of 0.14% with respect to the baseline scenario. However, the potential environmental impact is greater. Deforested hectares would be reduced by 2.45% and GHG emissions by 0.19%. Notwithstanding, EUDR due diligence costs may still prevent compliant production from entering the EU market, so the total impacts could be higher.

**Keywords**: Deforestation-free products, European Union, Argentina, exports, trade regulations.

**JEL**: C68, F13, F18, F4.

[1] Fundar e Instituto de Investigaciones Económicas (Facultad de Ciencias Económicas, Universidad Nacional de La Plata, Argentina). This paper is part of my thesis for obtaining the degree of PhD in Economics at the Universidad Nacional de La Plata. I am very grateful to Martin Cicowiez for his helpful and insightful comments. Email: delavegapc@gmail.com.

# 1. Introduction

Deforestation is one of the main causes of climate change and biodiversity loss globally (IPBES, 2019; IPCC, 2023) and the conversion of forest areas to produce commodities has been identified as one of its most important determinants (Pendrill et al. 2022)[2]. In this context, restrictions on the consumption of products produced on deforested land emerge as a way to combat this phenomenon globally and fight against climate change.

As of January 2026, the European Union (EU)[3] will require a traceability system on certain raw materials and derived products identified as drivers of global deforestation.[4] This policy, known as the European Union Deforestation Regulation (EUDR), is part of a broad package of measures included in the Green Deal, where the EU outlined the guidelines to achieve carbon neutrality by 2050 (European Parliament, 2023; CEI, 2023).[5]

The EUDR covers cattle, cocoa, coffee, oil palm, rubber, soybeans and timber and their products such as meat, chocolate, leather and paper (European Parliament, 2023). Such products may only be exported to the EU if they are supported by a due diligence process demonstrating that they have been produced following the legislation of the producing country and are "deforestation-free". The EUDR defines "deforestation-free" as those raw materials and products produced on land that did not undergo deforestation after December 31, 2020, and timber that has been harvested from forests without inducing forest degradation after December 31, 2020 (Drost et al., 2022; Stam, 2023; Calvo et al., 2024).[6] The EU Regulation then prohibits even deforestation that was legally permitted under domestic regulations. The EUDR adopts the FAO definition of forest: *land with a tree canopy cover of more than 10 percent and area of more than 0.5 ha. The trees should be able to reach a minimum height of 5 m at maturity in situ"* (FAO, 2018).

---

[2] The products mainly associated with deforestation are beef, forestry products, palm oil, cereals and soybeans.
[3] Currently composed of 27 (twenty-seven) countries: Austria, Belgium, Bulgaria, Croatia, Cyprus, Czech Republic, Denmark, Estonia, Finland, France, Germany, Greece, Hungary, Ireland, Italy, Latvia, Lithuania, Luxembourg, Malta, Netherlands, Poland, Portugal, Romania, Slovakia, Slovenia, Spain, Sweden and Slovakia.
[4] Initially scheduled to take effect in 2025, its application has been postponed to 2026 (see here and here).
[5] Carbon neutrality implies achieving a net result of zero greenhouse gas emissions, that is, emitting the same number of gases that are absorbed in other ways.
[6] Deforestation refers to the removal of forests to use the land for productive activities. Meanwhile, forest degradation is a gradual process that decreases forest biomass, changing its composition or reducing the quality of its soil. Therefore, forest degradation is more difficult to measure and monitor and statistics are scarce (European Commission, 2021).

Argentina is a middle-income country in Latin America that is significantly exposed to the EUDR. The EU is its second largest export destination, and we estimate that the EUDR would cover around 5909.66 million US dollars in exported value, but only 2.84% is not compliant with the EUDR, being soy and cattle the most affected production chains. Notwithstanding, EUDR due diligence costs may still prevent compliant production from entering the EU market, so the total impacts could be higher.

In addition, deforestation in Argentina has been significant in recent years: the country has lost 2% of its native forest cover between 2018 and 2022, representing more than one million hectares. According to Argentina's Native Forest Monitor, which uses the same definition of forest as the EUDR, this phenomenon has been particularly important in some provinces with high forest stocks such as Chaco, Formosa, La Pampa, Salta, and Santiago del Estero (see Figure 1). According to data provided by MapBiomas, between 2010 and 2020, 3 percent of natural forest was converted to agricultural use, and 1.25 percent between 2021 and 2022. In those provinces with high deforestation rates, increased land use has been mainly associated with crop production and livestock activities (Figure 2).

**Figure 1. Native Forest Stock in 2017 and cumulative deforestation between 2018 and 2022 (million hectares)**

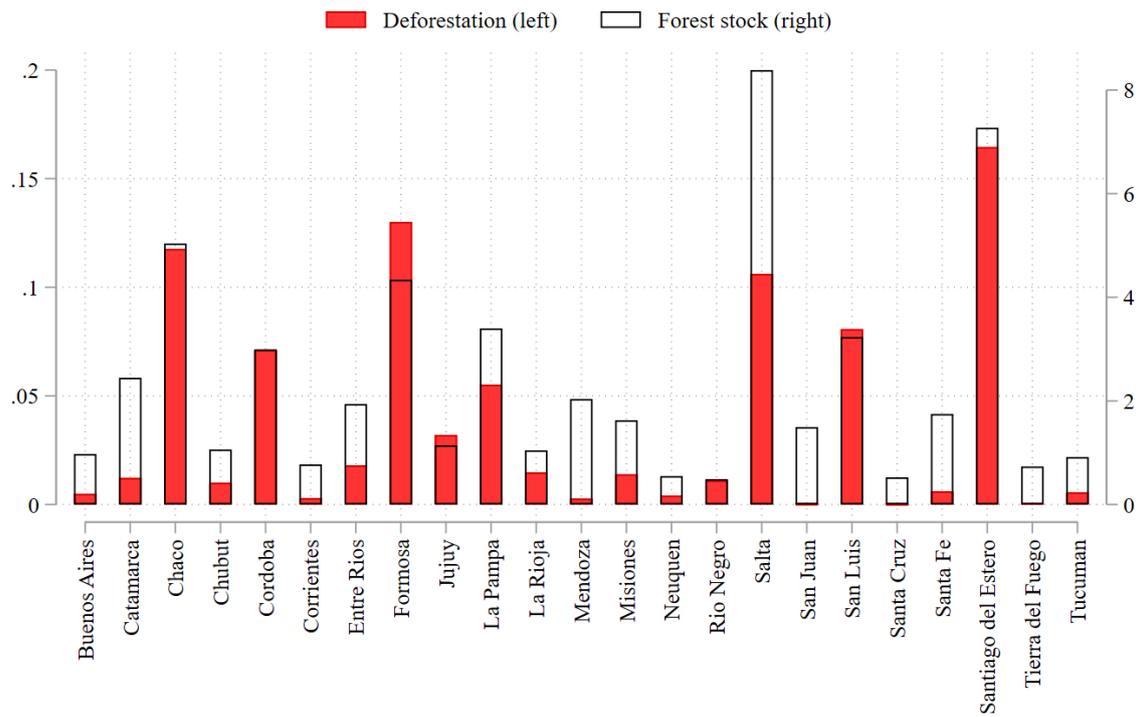

Own elaboration based on INDEC's National Forest Inventory and Native Forest Monitor.

**Figure 2. Productive Land Use (million hectares), 2010-2022**

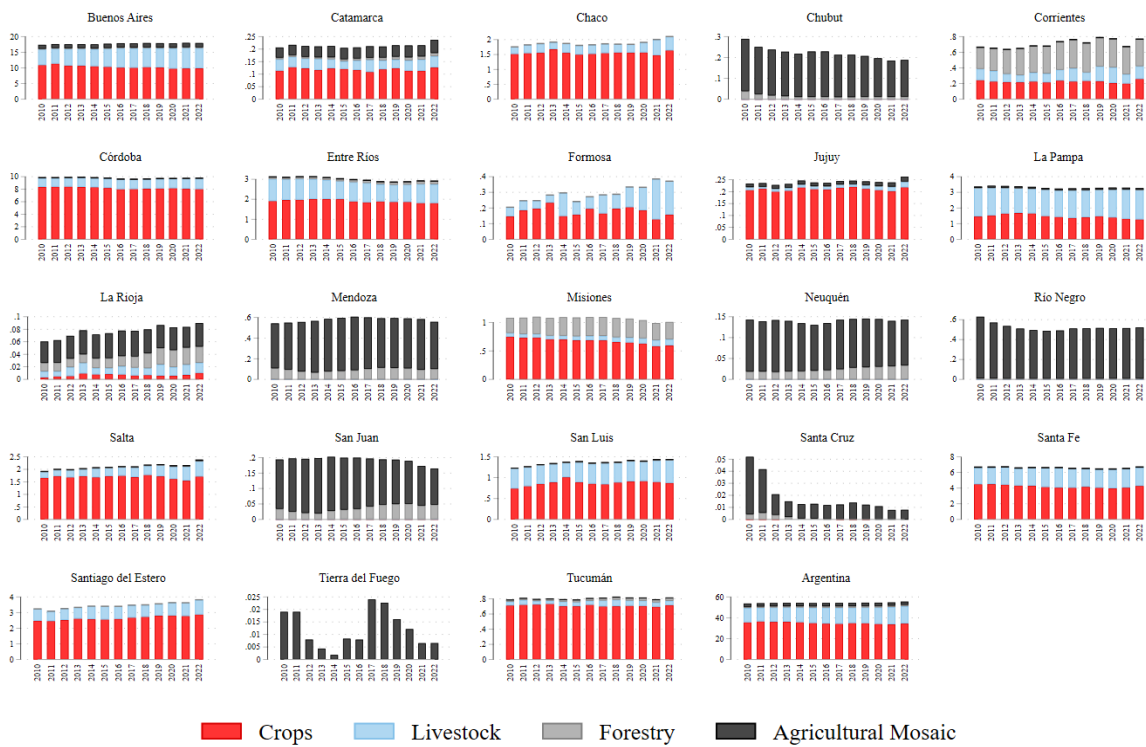

Own elaboration based on MapBiomas.

We use a computable general equilibrium (CGE) model to analyze the economic impact of the EUDR in Argentina. In order to introduce the EUDR shock in the model, we combine MapBiomas with agricultural census data to differentiate between compliant vs non-compliant production. In other words, we estimate the share of production that uses land that has been deforested after 2020 and thus is not compliant with the EUDR. The results suggest that the potential macroeconomic impacts are limited. As a consequence of the EUDR, between 2025 and 2030, GDP would be reduced by an average of 0.14% with respect to the baseline scenario. However, of greater magnitude is the potential environmental impact. Deforested hectares would be reduced by 2.45% and GHG emissions by 0.19%. In this way, this article contributes to the literature on the implications of unilateral climate-related trade measures, such as the EUDR, in developing countries, with novel evidence on the macroeconomic, sectoral, and environmental impacts in Argentina.

The remainder of this paper is as follows. In the next section, a literature review is carried out. Section 3 describes the CGE model used to simulate the impact of the EUDR on the Argentinean economy, which results are presented in Section 4. Section 5 performs a sensitivity analysis to evaluate the robustness of the results. Conclusions are included in Section 6.

## 2. Literature Review

This paper connects with several strands of the literature analyzing the consequences of domestic policies and international trade regulations that aim to reduce deforestation. Most of these papers use general equilibrium models but studies that use alternative modeling tools, including partial equilibrium models, will be also reviewed below. CGE models—both country models and global models such as GTAP—are widely used to analyze international trade policies with anti-deforestation objectives (Taheripour et al., 2019; Busch et al., 2022). Their main strength lies in capturing economy-wide linkages, factor reallocation, the macroeconomic constraints under which an economy usually operates, and leakage effects that partial equilibrium approaches cannot. In addition, they allow consistent integration of land-use change into global markets . Alternative modelling approaches—such as sector-specific partial equilibrium models (Jafari et al., 2017) or choice-based trade models (Hsiao, 2021)—provide finer sectoral or behavioural detail such as non-competitive structure in parts of the value-chain (Porcher and Marek, 2022; Dominguez-Iino, 2023), but omit cross-sectoral feedbacks and indirect effects.

Many of these studies focus on the case of palm oil, which has become the most widely consumed vegetable oil globally and whose demand has been largely met through the expansion of cultivated areas at the expense of tropical forests in South Asia. The evidence suggests that demand-side measures similar to the EUDR, including its more related precedent, the EU Timber Regulation[7], can modestly reduce deforestation in targeted countries, but leakage to other regions and commodities can offset part of the gains (Bosello et al., 2013; Taheripour et al., 2019; Busch et al., 2022). From a modelling perspective, these demand-side restrictions are typically implemented as a tariff or a reduction in the international price paid by the trade partner that imposes the restriction (Jafari et al., 2017; Taheripour et al., 2019; Busch et al., 2022). In addition, some papers explicitly differentiate two varieties of the targeted products depending on whether they are compliant or not (Busch et al., 2022). Key lessons are that the impacts depend on key land-supply elasticities, the share of targeted products in deforestation, and the ability to prevent leakage, i.e., the effectiveness of such demand-side measures increases with coordinated global action.

---

[7] The EU Timber Regulation is policy that prohibited the placing of illegally harvested timber and timber products on the EU market. According to the European Commision (2021), is the most comparable system to the EUDR.

The spirit of the EUDR also relates to several existing mechanisms for combating deforestation. These include, for example, zero deforestation commitments (ZDCs) made by the private sector to reduce or eliminate deforestation from their supply chains, and incentives provided under the Reduce Emissions from Deforestation and Forest Degradation mechanism (REDD), which provides for the transfer of resources to developing countries to compensate them for their efforts to reduce deforestation. The literature finds that, under full implementation and compliance, ZDCs can significantly curb forest loss but rely on perfect enforcement and global participation to avoid leakage (Mosnier et al., 2017; Leijten et al. 2023). Similarly, REDD mechanisms can contribute to protecting carbon-rich forest areas in developing countries to prevent them from being converted to productive land with limited GDP impacts (Overmars et al., 2014; Tabeau et al., 2015; Leitão et al, 2017; Francisco and Gurgel, 2020). In such models, ZDCs and REDD mechanisms are typically modeled as a reduction in the supply of land available for expansion by moving the asymptote of a stylized land supply curve.

Although recent, the study of the consequences of the measures contemplated in the EU's Green Deal has received considerable attention, especially the Carbon Border Adjustment Mechanism (CBAM). Both partial and general equilibrium models (UNCTAD, 2021; Chepeliev, 2021; Korpar et al., 2023; Michelena, 2023) find modest trade and GHG emissions effects, with distributional impacts depending on exporters' carbon intensity.

For the EUDR itself, quantitative evidence is still scarce. Existing CGE studies for Brazil (Stam, 2023), Argentina (Calvo et al., 2024), and Indonesia (Drost et al., 2022) suggest small aggregate GDP losses but heterogeneous sectoral and regional impacts, with limited GHG emissions reductions when supply is inelastic or the targeted product accounts for a small share of deforestation. From a modelling point of view, Stam (2023) focuses on the soy sector and introduces the EUDR shock as an increase in the production tax rate (whose revenue is not recycled for a specific use). The incremental cost is determined at the regional level by multiplying the share of soybean exports to the EU by 6%, which is the cost increase that emerged from a pilot test. Calvo et al. (2024) assess the short-term impact of the EUDR in Argentina, assuming that the country cannot adapt and all products listed in the regulation cease to be exported to the EU.

An important aspect of the EUDR is the cost related to the due diligence process, which depends on the size and complexity of the value chain. The European Commission (2021) suggests that the most comparable system to the EUDR is that associated with the EU Timber Regulation, whose compliance required set-up costs of between US$5,000 and US$90,000 per importer and recurrent costs of between 0.29 and 4.3% of the value imported. Using these percentages, the report estimates that total annual compliance costs for importing companies can be between 175 and 2,616 million euros per year (European Commission, 2021). Drost et al. (2022) focus on the Indonesian palm oil sector and estimate that EUDR compliance costs are relatively low and could be between 2.5 and 3.5% of the value of Indonesia's palm oil exports to the EU.

## 3. Methodology

This section describes the CGE model that will be used to analyze the isolated effect of the EUDR in Argentina. It is a dynamic-recursive CGE model, which is a modified version of the IEEM model of the Inter-American Development Bank (Banerjee and Cicowiez, 2019; Banerjee and Cicowiez, 2020; IDB, 2021). A small open economy is considered, where producers and consumers maximize profits and utility, respectively, in competitive markets.

Each productive sector is represented by a profit-maximizing activity. The value-added production technology is assumed to be represented by a CES-type function (Constant Elasticity of Substitution) combining capital, labor, and land, while intermediate inputs are used in fixed proportions. Regarding factor markets, capital is assumed to be fully employed and sector-specific, while labor supply is exogenous and labor is perfectly mobile across sectors. Additionally, rigidities are introduced using a wage curve, which allows to account for the negative empirical relationship between the wage level and the unemployment rate (Blanchflower and Oswald, 1994). Then, the equilibrium in the labor market is determined by the intersection of labor demand and the wage curve. The modeling of the land factor is described in detail in Section 3.2.

In terms of institutions, we model households disaggregated by decile of per capita family income that receive income from the productive factors they own and transfers from the government and the rest of the world. This income is used to pay direct taxes, to save, to make transfers to other institutions, and to consume goods. Private consumption demand is derived from the maximization of a Cobb-Douglas utility function. The government collects taxes on households, factors, activity, sales, and foreign trade, and receives transfers from the rest of the world. It then uses this revenue to purchase goods for consumption, invest, make transfers to households, and save. As already mentioned, a small economy is modeled, so that international prices are exogenous. As usual in the literature, following Armington (1969), imperfect substitution is assumed between goods that differ according to their origin, so that the demand for imports arises from a CES (Constant Elasticity of Substitution) function that combines domestic and imported goods. Meanwhile, the supply of exports is modeled from a CET (Constant Elasticity of Transformation) function, which reflects the fact that producers decide to allocate their production to the domestic market or to export it depending on relative prices.

The model is recursive-dynamic so that agents' expectations are myopic. The sources of dynamics are capital accumulation, labor force growth, and productivity change. Investment modifies the next period's capital stock, labor supply grows exogenously according to population projections, and land supply grows as a function of deforestation, as explained in Section 3.2.

The model requires the specification of closure rules for three macroeconomic balances: government, private savings and investment, and the balance of payments. In this regard, all simulations assume that: i) the government budget is balanced by changes in real domestic financing; ii) private investment is endogenously determined by the level of savings; and iii) savings from the rest of the world are exogenous (measured in foreign currency), so that the real exchange rate varies endogenously to match inflows and outflows of foreign exchange.

### 3.1 EUDR modeling

The EUDR prohibits the export to the EU of raw materials and products produced on deforested land after the cut-off date of December 2020.[8] Being a retrospective measure, today some decisions are sunk and the land factor can be divided into two types: deforested after December 2020 (non-compliant) and not deforested or deforested before January 2021 (compliant). Production can be divided into analogous terms: that produced on non-deforested land will face an additional cost associated with the EUDR due diligence process if exported to the EU, while that produced on deforested land could only be sold to other destinations or in the domestic market (if there is a domestic demand for such products).

---

[8] We abstract from the ban on forest degradation, which is a more difficult process to measure and for which statistics are scarce (European Commission, 2021).

The objective then is to disaggregate the land factor into deforested and non-deforested land and land-demanding activities (crops, livestock, and forestry) and the products they produce (and consume) from deforested and non-deforested land.[9] Finally, we will also disaggregate industrial products that, although not produced by land-demanding activities, are produced by activities that use inputs whose production demands land (they demand land "indirectly"). The latter is important because the EUDR will require traceability throughout the chain.

We use the data provided by MapBiomas to estimate the share of land used in agricultural, livestock and forestry activities that has been deforested after 2020 and thus is not compliant with the EUDR. MapBiomas is a collaborative network of NGOs, universities and technology companies dedicated to mapping land cover and land use changes in South America and Indonesia. They use the Google Earth Engine platform and artificial intelligence to perform a pixel-by-pixel classification of land cover and land use. The procedure is as follows:

First, we identify forests in the MapBiomas data. MapBiomas identify "natural wooded vegetation" that includes closed woodland, open woodland, sparse woodland and flooded woodland. However, the application of the FAO's definition of forest varies by region/province (MapBiomas, 2022). Then, we calculate the hectares of forest land that was converted to productive land after 2020 and the share of EUDR non-compliant land by activity as follows:

$$noncompliantlandt_{i,r} = \frac{D_{i,r,2021} + D_{i,r,2022}}{L_{i,r,2022}}$$

where $noncompliantlandt_{i,r}$ is the share of non-compliant land used in activity $i = \{crop, livestock, forestry\}$ in region $r$; $D_{i,r,t}$ is the amount of hectares of forest land that was converted to land used by activity $i$ in region $r$ in year $t$; and $L_{i,r,2022}$ is the amount of hectares of land used by activity $i$ in region $r$ in year 2022, the last year with information available. Unfortunately, MapBiomas provides data only at the activity level (crop, livestock and forestry) so we combine this information with the 2018 Agricultural Census Data. The Census provides a further disaggregation of land used at the crop-level. Thus, the share of non-compliant land varies by crop across provinces but is constant for crops within a province.

---
[9] A similar approach is used by Busch et al (2022), see Section 2.

Finally, we use the estimated share of non-compliant land to split production (and land factor use) of land-demanding activities (crops, livestock, and forestry) into compliant and non-compliant with the EUDR. Finally, industrial products that demand land "indirectly" are disaggregated using the main raw material distribution factors. For example, soybeans in the case of industrial soybean oil production.

## 3.2 Land modelling

Land supply curves with constant price elasticity are introduced to endogenize the deforestation path and model the growth of land supply, as follows:

$$QDEFOR_{fland-defor,t} = QFS^{00}_{fland-defor} \left[ \left( \frac{\frac{WFAVG_{fland-defor,t}}{CPI_t}}{\frac{WFAVG^{00}_{fland-defor}}{CPI^{00}}} \right)^{\mu_{fland-defor}} - 1 \right] \quad (1)$$

where $fland - defor$={Crop land (non-compliant); Livestock land (non-compliant); Forestry land (non-compliant)}; $QDEFOR_{f,t}$ is the is the amount of land deforested for use as a productive factor $f$ in the period $t$; $QFS^{00}_f$ is the supply of land factor $f$ in the base year; $WFAVG_{f,t}$ y $WFAVG^{00}_f$ are the average remuneration of the land factor $f$ in period $t$ and in the base year, respectively; $CPI_t$ y $CPI^{00}$ is the consumer price index in period $t$ and in the base year, respectively; and $\mu_f$ is the supply price-elasticity of land factor $f$. Thus, equation (1) models the induced deforestation associated with the increase in the supply of the $fland - defor$ land factor as a function of the increase in its relative remuneration.

In the following period, this additional land is added to the initial offer of the next period:

$$QFINIT_{fland-defor,t} = QFS_{fland-defor,t-1} + QDEFOR_{fland-defor,t-1} \quad (2)$$

where $QFINIT_{f,t}$ is factor $f$ initial supply in period $t$; and $QFS_{f,t}$ is land factor f supply in period $t$.

The analogous equation for the non-deforested land factors is:

$$QFINIT_{fland-nodefor,t} = QFS_{fland-nodefor,t-1} \tag{3}$$

where $fland - nodefor$={Crop land (compliant); Livestock land (compliant); Forestry land (compliant)}. In both cases ($fland - nodefor$ and $fland - defor$), migration between land uses (crops, livestock or forestry) is then allowed according to their relative profitability (BID, 2021).[10] No migration of the land factor from uses using deforested land to non-deforested land and vice versa is allowed.

Total deforestation during the period $t$, $QDEFORTOT_t$, is:

$$QDEFORTOT_t = \sum_{fland-defor} QDEFOR_{fland-defor,t} \tag{4}$$

which decreases the non-productive forest area, $QLAND_{fornprod,t}$, in the following period:

$$QLAND_{fornprod,t} = QLAND_{fornprod,t-1} - QDEFORTOT_{t-1} \tag{5}$$

In summary, while the supply of "non-deforested" land remains constant, deforestation reduces the non-productive forest area and increases the initial supply of the "deforested" type of land in the next period. Finally, unemployment of both land types is allowed for by functional forms analogous to the labor factor wage curves.

### 3.3 Calibration

The main source of information used to calibrate a CGE model is a Social Accounting Matrix (SAM), an accounting record of all economic transactions in an economy in a given period. Each account is represented by a row (income) and a column (expenditure). The row sum of each account is matched to the column sum of the same account, thus respecting the budgetary constraints of each agent and the sectoral and macroeconomic supply and demand balances.

We construct the SAM for Argentina 2019 following the methodology of Banerjee and Cicowiez (2021) and using the Supply and Use Tables provided by INDEC. Additionally, the rest of the world account is disaggregated into four main blocks using data on international trade flows: the EU and the Rest of the trading partners.

---

[10] The use of this migration module has the advantage over other ways of modeling land use change such as that using CET functions, in which they fail to maintain the balance in the physical land units (Taheripour et al., 2020).

In order to translate the list of products covered by the EUDR to the Argentine data, we proceeded as follows. First, we built a correspondence between that list and the Harmonized System (HS) classification. Then, we merge the HS list with the Central Product Classification (CPC) that is used in Argentina's Supply and Use Tables. Unfortunately, this last is more aggregated, thus we end up with a list of products that include at least one product covered by the EUDR. For example, soy and derived products are included in "Oilseeds and oil fruit"; and "Residues from the extraction of vegetable fats; oilseed meals". Although this may lead to imposing the impact on production that is not covered by the EUDR, there are no alternative data to the INDEC supply and use tables; that is, they are the most disaggregated data available for Argentina.

The SAM information is complemented with elasticities and other behavioral parameters from the literature. In particular, our model requires values for i) the wage-unemployment elasticity; ii) the elasticities of substitution between domestic purchases and imports (Armington); iii) the elasticities of transformation between domestic sales and exports (CET); iv) the elasticity of transformation of exports between destinations; v) the price elasticity of land supply; vi) the unemployment elasticity of land. For this work, these inputs are obtained from the resources publicly provided by the Open-IEEM Project of the Inter-American Development Bank (IDB, 2021). The wage-unemployment elasticity is -0.1, and the Armington and CET elasticities are both in the range of 0.9-2 (see Table A1 in Appendix A). The price elasticity of land supply is calibrated so that the average deforestation rate in the baseline scenario is equal to the historical average between 2000 and 2019, which was 0.8% according to FAO.

The labor unemployment rate is obtained from ILOSTAT. Land endowments are obtained from FAOSTAT and the 2018 National Agricultural Census. Land unemployment rates are calculated using uncultivated suitable areas from the 2018 National Agricultural Census. Emissions of polluting gases, measured in million tons of carbon dioxide equivalent (MtCO2e) are obtained from the National Greenhouse Gas Inventory. A particular distinction is made between emissions associated with agriculture, livestock, forestry, and land use change (AFOLU). In the latter case, emissions are assigned to the corresponding activities (crops, livestock, or forestry) or to non-productive forest land (which can sequester carbon). In the case of "non-AFOLU" emissions, inventory emissions are allocated to product consumption (e.g., oil) and distributed among emitters (domestic activities or institutions) according to their consumption (intermediate or final) as derived from the SAM.

Population projections are obtained from the United Nations, while the INDEC's National Household Income Expenditure Survey is used to disaggregate households by deciles of per capita family income. GDP growth projections are obtained from the IMF WEO (April 2024).

**3.4 Scenarios**

The baseline is a business-as-usual scenario that simulates the Argentine economy between 2019 and 2030, assuming the absence of the EUDR. An exogenous path for GDP growth is assumed based on IMF projections, while total factor productivity is determined endogenously. The resulting productivity path is then used as exogenous in the counterfactual scenario. The shock scenario simulates the evolution of the economy in the same period but introduces the EUDR in two ways. On the one hand, as a reduction in the international price paid by the EU for non-compliant products that were produced on deforested land or consumed intermediate inputs produced on deforested land, so that their quantity exported to the EU is approximately zero.[11] On the other hand, the export to the EU of compliant products produced on non-deforested land will face an additional cost associated with the EUDR due diligence process. As reviewed in Section 2, there is significant uncertainty regarding the costs of this process. Based on estimates by the European Commission (2021), Drost et al. (2022) and Stam (2023), this incremental cost is introduced as an international price reduction of 6%.[12]

Table 1 describes the magnitude of the EUDR shock in Argentina through the share of EU exports, the share of exports in the total demand for each product (i.e., the importance of the domestic market), and the exported value to the EU in US dollars. The direct exposure is high for products such as waste products, oilseeds products, leather products, animal and meat products, where the EU accounts for a large share in export destinations and exports are an important part of the total demand. With respect to the non-compliant production, measured in US dollars, the shock is particularly significant for other animal products (68.25 million USD), meat products (27.66 million USD), oilseed meals and its residues (27.58 million USD), oilseeds and fruits (15.88 million USD) and waste products (15.63 million USD). The total value exported to the EU of non-compliant products is 167.62 million USD. Meanwhile, for the compliant production, the exported value to the EU is particularly high for oilseed meals and its residues (2670.85 million USD), oilseeds and fruits (1537.64 million USD), meat products (937.2 million USD), waste products (190.97 million USD), leather products (121.44 million USD), and animal and vegetable oils and fats (119.3 million USD). The total value exported to the EU of compliant products is 5742.04 million USD. In sum, the EUDR would cover around 6 billion US dollars in exported value, but only 2.84% is not compliant with the EUDR, being soy and cattle the most affected production chains.

---

[11] This procedure is similar to that used by Busch et al (2022), who introduce the ban on the "high deforestation" palm oil variety as a tariff in Europe high enough that imports are reduced by 99% (see Section 2).

[12] While this reduces the incentive to export these products to the European Union, it does not generate an increase in the cost of production. Given the production and marketing structure modeled in the CGE model, there is no obvious way to introduce a cost at the production level that is linked to exporting to a particular destination, as the decision is made to sell production in the domestic market or export it at another level.

**Table 1. Direct EUDR exposure, by product**

| | % EU in exports | % Exports in demand | Exported Value to the EU (million USD) | |
|---|---|---|---|---|
| | | | Non-compliant | Compliant |
| Oilseeds and fruits | 35.74 | 26.07 | 15.88 | 1537.64 |
| Crops of drinkable plants and spices | 12.39 | 1.31 | 0.03 | 0.53 |
| Living animals | 9.27 | 0.29 | 0.09 | 2.98 |
| Other animal products | 28.31 | 4.78 | 68.25 | 2.01 |
| Timber and other forestry products | 52.55 | 6.50 | 0.99 | 19.25 |
| Meat and meat products | 23.04 | 24.84 | 27.66 | 937.20 |
| Animal and vegetable oils and fats | 2.55 | 35.10 | 1.23 | 119.30 |
| Residues from the extraction of vegetable fats; oilseed meals | 29.92 | 85.05 | 27.58 | 2670.85 |
| Cocoa, chocolate and confectionery | 1.39 | 4.19 | 0.15 | 1.88 |
| Food products n.e.c. | 1.55 | 6.58 | 0.08 | 7.18 |
| Leather and other leather goods | 21.84 | 46.10 | 3.58 | 121.44 |
| Wood | 0.75 | 10.20 | 0.05 | 1.00 |
| Carpentry works and parts for construction | 1.16 | 0.41 | 0.00 | 0.02 |
| Wooden boxes and containers, cooperage products | 8.22 | 0.26 | 0.00 | 0.05 |
| Other wood, cork and braiding materials products | 15.60 | 0.74 | 0.05 | 1.06 |
| Pulp, paper and cardboard | 0.49 | 5.24 | 0.10 | 2.02 |
| Printed products (except advertising) | 10.79 | 3.14 | 0.19 | 3.69 |
| Newspapers, magazines and periodicals | 1.14 | 0.47 | 0.00 | 0.06 |
| Advertising material and other printed matter | 3.21 | 0.74 | 0.01 | 0.21 |
| Record books, stationery, paper or cardboard | 0.99 | 0.12 | 0.00 | 0.01 |
| Basic organic chemicals | 23.30 | 8.10 | 4.45 | 90.31 |
| Miscellaneous basic chemicals | 16.13 | 48.52 | 1.26 | 25.65 |
| Tires and other rubber products | 2.35 | 8.50 | 0.31 | 6.10 |
| Furniture | 2.06 | 0.96 | 0.03 | 0.62 |
| Prefabricated buildings | 0.01 | 5.72 | 0.00 | 0.00 |
| Waste or scrap | 46.43 | 30.09 | 15.63 | 190.97 |
| Total | | | 167.62 | 5742.04 |

Own elaboration.

## 4. Results of the Counterfactual Scenarios

The EUDR shock reduces the price the EU pays for the products covered by the regulation. In the case of those produced on deforested land, i.e., non-compliant, the price reduction is such that the quantity exported becomes practically zero, while in the case of those produced on non-deforested land, and thus compliant, it is a relatively minor price reduction, which allows for continued exports. In both cases, there is an incentive to divert production to other destinations or the domestic market (if there is domestic demand for such products). However, for reasonable values of the elasticities (of transformation between domestic sales and exports, and between export destinations), it is to be expected that the diversion of sales will not be able to compensate for the lost sales to the EU, so a fall in exports and production would be evident. In aggregate terms, the EUDR is expected to induce a reduction in non-compliant production and thus deforestation compared to the baseline scenario.

Table 2 shows the results in terms of production, domestic sales, exports and imports, for the products covered by the measure and which are produced on deforested land, so are non-compliant with the EUDR. For all these products, production falls, but particularly for timber and other forestry products, crops of drinkable plants and spices, and animal products, which face substantial reductions in exports due to a high share of the EU in exports as discussed in the previous section. However, the domestic market helps to cushion the impact, and declines in production are much smaller than reductions in exports.. In cases where production is more biased towards the domestic market (see column 2 in Table 1), either for intermediate or final consumption, domestic sales contribute to absorbing the shock. This is the case, for example, of leather and its derivatives, chemical products, meat and meat products, and oilseeds and oleaginous fruits. Among these, there are also cases in which exports to other destinations also marginally grow, although they account for a small part of the demand. Examples of the latter are prefabricated buildings, carpentry pieces, advertising material, newspapers and magazines, and record books, among others.

**Table 2. Results by product, selected cases (average % deviation with respect to baseline, 2025-2030)**

| | Production | Domestic Sales | Exports | Imports |
|---|---|---|---|---|
| Oilseeds and fruits (non-compliant) | -0.68 | 0.94 | -14.62 | -8.02 |
| Crops of drinkable plants and spices (non-compliant) | -0.71 | -0.60 | -10.18 | 1.73 |
| Living animals (non-compliant) | -0.58 | -0.55 | -17.80 | 13.61 |
| Other animal products (non-compliant) | -0.43 | -0.01 | -15.65 | -1.87 |
| Timber and other forestry products (non-compliant) | -1.38 | 0.37 | -33.57 | -2.11 |
| Meat and meat products (non-compliant) | -0.10 | 1.21 | -9.94 | -1.69 |
| Animal and vegetable oils and fats (non-compliant) | -0.10 | -0.07 | -0.80 | -0.65 |
| Residues from the extraction of vegetable fats; oilseed meals (non-compliant) | -0.10 | -0.10 | -8.26 | 0.00 |
| Cocoa, chocolate and confectionery (non-compliant) | -0.10 | -0.12 | -0.02 | -0.43 |
| Food products n.e.c. (non-compliant) | -0.10 | -0.12 | -0.28 | -0.52 |
| Leather and other leather goods (non-compliant) | -0.10 | 2.21 | -7.56 | -1.93 |
| Wood (non-compliant) | -0.10 | -0.11 | -0.23 | -0.21 |
| Carpentry works and parts for construction (non-compliant) | -0.10 | -0.10 | 0.21 | -0.65 |
| Wooden boxes and containers, cooperage products (non-compliant) | -0.10 | -0.10 | -4.96 | -0.31 |
| Other wood, cork and braiding materials products (non-compliant) | -0.10 | -0.06 | -9.01 | -0.82 |
| Pulp, paper and cardboard (non-compliant) | -0.10 | -0.12 | 0.03 | -0.32 |
| Printed products (except advertising) (non-compliant) | -0.10 | 0.00 | -5.57 | -0.50 |
| Newspapers, magazines and periodicals (non-compliant) | -0.10 | -0.11 | 0.14 | -0.41 |
| Advertising material and other printed matter (non-compliant) | -0.10 | -0.10 | -0.62 | -0.84 |
| Record books, stationery, paper or cardboard (non-compliant) | -0.10 | -0.10 | 0.08 | -0.71 |
| Basic organic chemicals (non-compliant) | -0.10 | 1.02 | -12.15 | -1.11 |
| Miscellaneous basic chemicals (non-compliant) | -0.10 | 1.32 | -5.27 | -2.44 |
| Tires and other rubber products (non-compliant) | -0.10 | -0.05 | -1.08 | -0.34 |
| Furniture (non-compliant) | -0.10 | -0.11 | -0.45 | -0.61 |
| Prefabricated buildings (non-compliant) | -0.10 | -0.20 | 0.52 | -0.80 |
| Waste or scrap (non-compliant) | -0.10 | 0.76 | -21.16 | -11.92 |

Own elaboration.

These results are reflected at the level of productive sectors (Figure 3). Those that suffer the most are the non-compliant activities, whose value-added falls by 1.59% in the case of the forestry sector, 0.71% for crop activities and 0.85% for the cattle sector. Compliant activities also fall but to a lesser extent. This translates into important effects on the land market, as shown in Figure 4, reducing the demand for non-compliant land and therefore deforestation. As for the rest of the sectors, the adjustment of the trade balance associated with the EUDR shock depreciates the exchange rate, boosting exports but also making imported inputs more expensive, which are relevant for sectors such as mining, manufacturing, and services. Overall, value added fell 0.12%.

**Figure 3. Results Value-added, by activity (average % deviation from baseline, 2025-2030)**

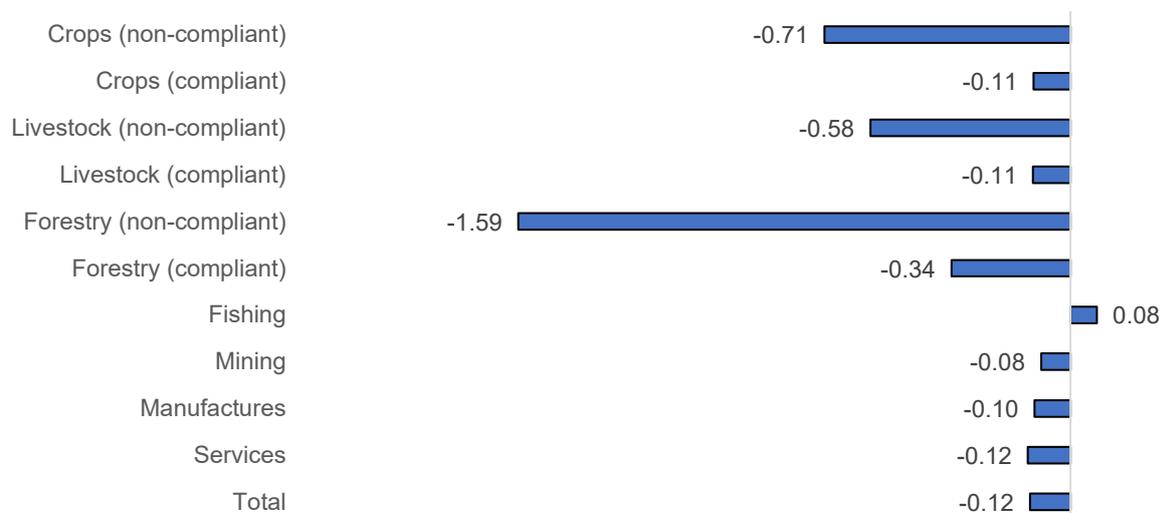

Own elaboration.

**Figure 4. Land Use Results (average % deviation from baseline, 2025-2030)**

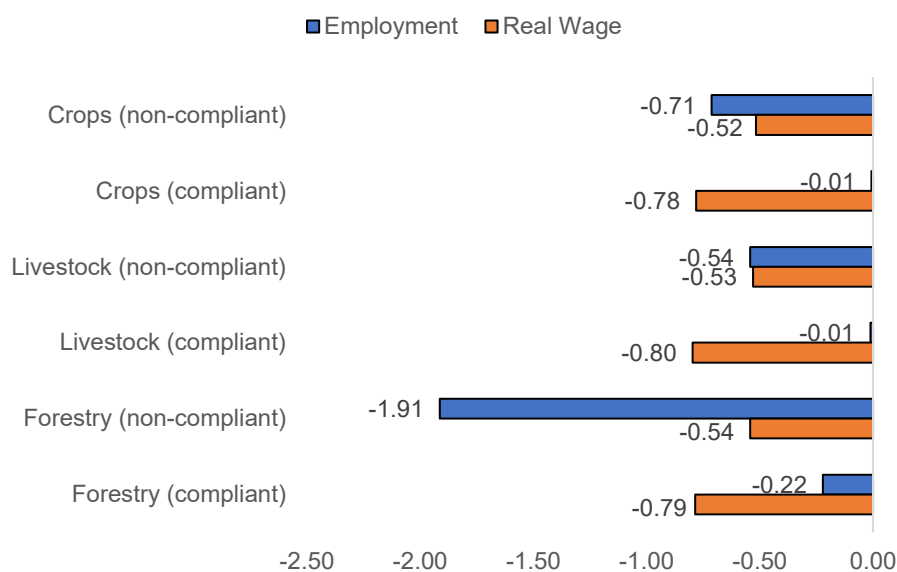

Own elaboration.

The sectoral impacts will be transmitted to the rest of the economy, for example, by inducing a depreciating tendency in the exchange rate, reducing the demand for and remuneration of productive factors, the income of the owners of such factors, government revenues, etc. Figure 5 summarizes the main macroeconomic, labor market, and environmental results. Given that the sectors most affected by the EUDR are not labor-intensive and account for a low proportion of labor demand, the impacts on the labor market are limited. Real wages fall by 0.36%, which explains a reduction in private consumption of 0.17%. As expected, the reduction in exports to the EU is not fully offset by sales to other destinations, so total exports are reduced by an average of 0.12%. This is associated with a real exchange rate 0.36% higher than in the base scenario, which adds up to a negative shock for those sectors that import inputs. In effect, imports fall 0.47% and GDP is reduced by 0.14%.

**Figure 5. Summary of macro, labor and environmental results (average deviation from baseline, 2025-2030)**

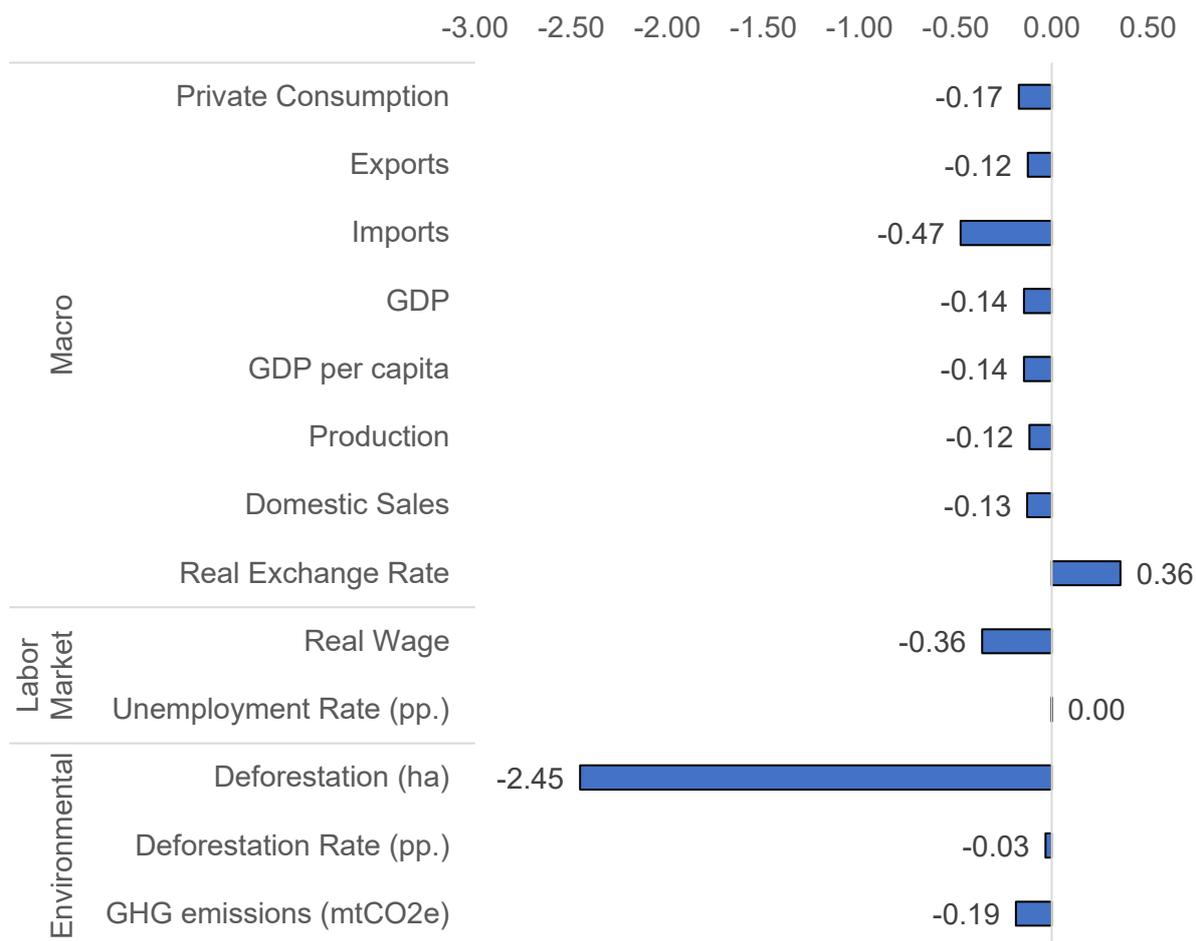

Own Elaboration. The figure shows the average percentage deviation (unless otherwise noted) from the baseline scenario (without EUDR) for key indicators.

As discussed above, the decrease in the demand for non-compliant land translates into a reduction in deforested hectares by an average of 2.45%, with the deforestation rate 0.03 percentage points lower. This represents almost 47 thousand fewer deforested hectares in the shock scenario during the period 2025-2030. Meanwhile, emissions of GHG gases are reduced by an average of 0.18% with respect to the base scenario, mainly because of a scale effect, whether due to decreases in activity or final or intermediate consumption (Figure 6). On the other hand, the composition effect refers to the share of the emitter among the consumers of emitting factors. It is positive, for example, for fishing activity, whose value-added expands, while it is negative for primary non-complaint activities, which contract (see Figure 3).

In sum, the macroeconomic impacts are modest, but some non-compliant sectors are highly impacted by the EUDR shock, such as soy, cattle, and timber production. The environmental impacts, on the other hand, are higher, mainly in terms of deforestation reduction.

**Figure 6. Results Pollutant gas emissions, by emitter (average % deviation from baseline, 2025-2030)**

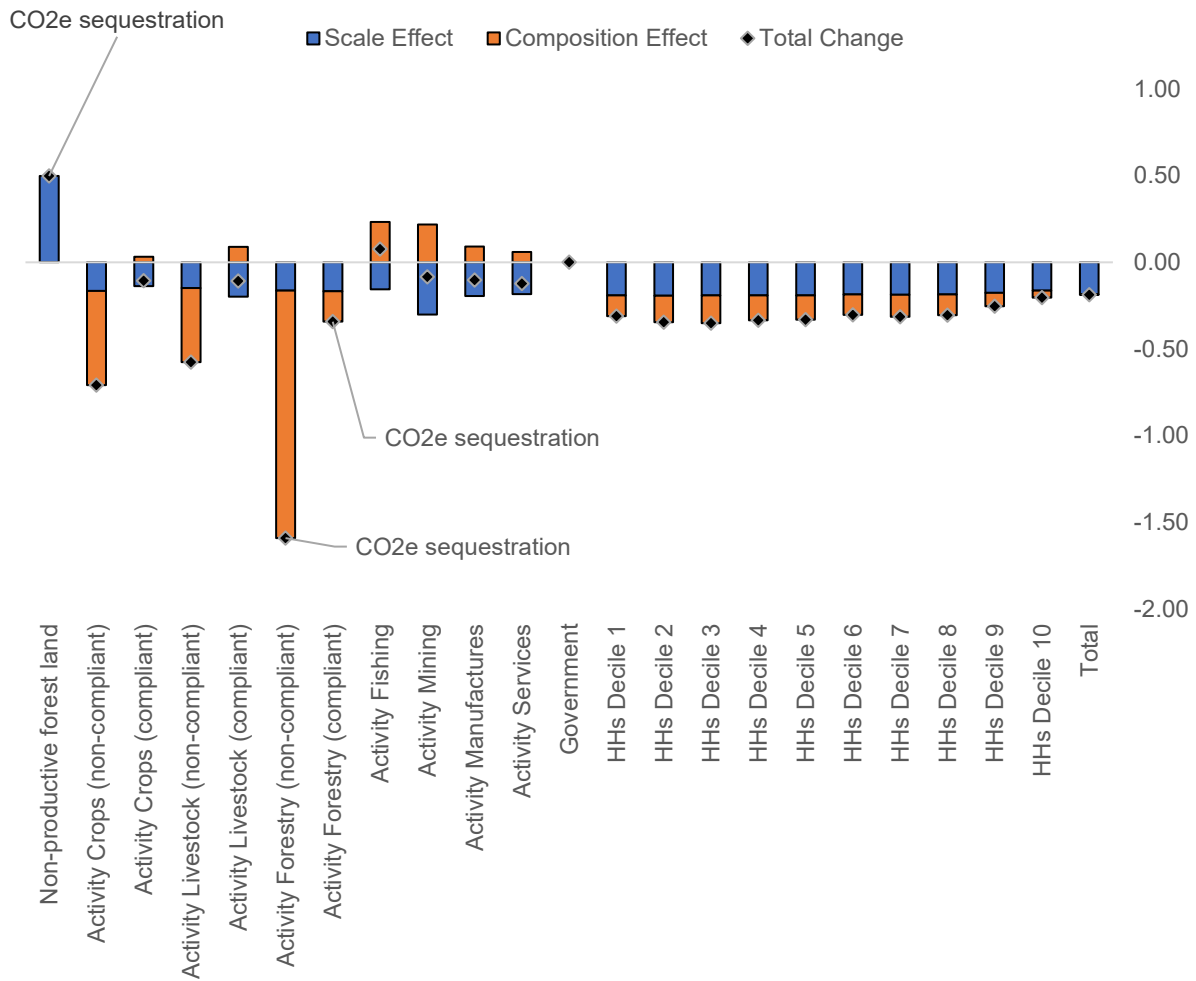

Own elaboration. The decomposition method is based on McMillan and Rodrik (2011). In cases involving an emitter that captures CO2, such as non-productive forest land, a positive total change represents an increase in sequestration (a greater reduction in emissions), and vice versa for a negative total change, as in this case for forestry activities.

## 5. Sensitivity Analysis

We assess the sensitivity of the results by modifying the assumptions regarding the magnitude of key parameters in our analysis. The elasticities to be modified are important to analyze the degree of adjustment of the production most exposed to the EUDR, which are agricultural activities and derived products. In this sense, the two important aspects of the problem are the changes in land use and thus deforestation, and the flexibility to modify the destination of sales either to the domestic market or to other export markets, which is usually called "leakage". For this reason, the sensitivity analysis is carried out by modifying: (i) the elasticity of land supply; (ii) the elasticity of the land wage curve; (iii) the elasticities of transformation between domestic sales and between exports; and (iv) the elasticities of transformation of exports between different destinations. While (i) and (ii) regulate changes in land use and the intensity of deforestation, (iii) and (iv) address the extent to which producers can circumvent the EUDR by selling their products to different markets. The value of the elasticities is modified in a wide range between +/-50%, as shown in Table 3.

**Table 3. Parameters for sensitivity analysis**

| ID | Description | EUDR Shock Scenario (Section 4) | Variation |
|---|---|---|---|
| S1 | Deforested Land Supply Curve (LO) | 0.06 | - 50% |
| S2 | Deforested Land Supply Curve (UP) | 0.06 | + 50% |
| S3 | Land Wage Curve (LO) | -0.4 | - 50% |
| S4 | Land Wage Curve (UP) | -0.4 | + 50% |
| S5 | Transformation between domestic sales and exports (LO) | Table A1 | - 50% |
| S6 | Transformation between domestic sales and exports (UP) | Table A1 | + 50% |
| S7 | Transformation between exports to different destinations (LO) | Table A1*2 | - 50% |
| S8 | Transformation between exports to different destinations (UP) | Table A1*2 | + 50% |

Own Elaboration.

The results are shown in Table 4, including a column with the results presented in Section 4, for comparison. In general terms, no major variations are found in terms of macroeconomic outcomes, value-added, and distributional impacts. However, there are important changes in environmental outcomes and land use changes in some simulations.

The lower the absolute value of the land supply elasticity (S1 vs. S2), the greater the reduction in deforested hectares and GHG emissions, which is associated with larger declines in land use. In the S1 scenario, for example, land use by non-compliant primary activities falls between 2 and 4% because of a less elastic land supply curve. Something similar occurs for the elasticity of the land wage curve (S3 vs S4), but in this case, in S3 is easier to use unemployed deforested land, so these activities are less dependent on deforestation.

Changes in the magnitudes of the elasticities of transformation between domestic sales and between exports, and the elasticities of transformation of exports between different destinations mainly generate changes in macroeconomic, sectoral, and distributional outcomes. The more difficult it is to transform exports into domestic sales (S5 vs. S6), the smaller the drop in total exports and the lower the real depreciation required to adjust the trade balance. However, domestic sales suffer more. On the contrary, the more difficult it is to divert exports to other destinations (S7 vs. S8) the larger is the drop in exports, the required adjustment in the real exchange rate, and the fall in GDP.

It is important to note that the intervals of the sensitivity analysis (+/- 50%) are really big. During the 2025-2030 period, for example, the cumulative number of deforested areas is 30% higher in the S2 scenario than in the base scenario (517 hectares more), while it is 26% lower in S1 (448 hectares more). Notwithstanding, while there is an uncertainty regarding these key parameters, the macro, sectoral, and labor results are robust and both qualitatively and quantitatively similar under very extreme assumptions regarding land use changes and leakage feasibility.

**Table 4. Sensitivity Analysis Results (average % deviation from baseline, 2025-2030)**

|  | EUDR | S1 | S2 | S3 | S4 | S5 | S6 | S7 | S8 |
|---|---|---|---|---|---|---|---|---|---|
| **Macroeconomic Results** | | | | | | | | | |
| GDP | -0.14 | -0.15 | -0.14 | -0.11 | -0.17 | -0.13 | -0.15 | -0.16 | -0.14 |
| Exports | -0.12 | -0.14 | -0.12 | -0.05 | -0.18 | 0.14 | -0.27 | -0.13 | -0.12 |
|    European Union | -7.02 | -7.02 | -7.02 | -6.96 | -7.06 | -3.62 | -9.91 | -4.07 | -9.70 |
|    Rest | 1.47 | 1.45 | 1.48 | 1.55 | 1.41 | 1.00 | 1.96 | 0.78 | 2.08 |
| Imports | -0.47 | -0.48 | -0.47 | -0.40 | -0.52 | -0.24 | -0.60 | -0.52 | -0.44 |
| Production | -0.12 | -0.12 | -0.11 | -0.08 | -0.14 | -0.12 | -0.11 | -0.13 | -0.11 |
| Domestic Sales | -0.13 | -0.13 | -0.12 | -0.09 | -0.15 | -0.16 | -0.11 | -0.14 | -0.12 |
| Real Exchange Rate | 0.36 | 0.36 | 0.36 | 0.30 | 0.40 | 0.23 | 0.41 | 0.39 | 0.33 |
| **Value-Added Results** | | | | | | | | | |
| Crops (non-compliant) | -0.71 | -1.74 | -0.05 | -0.49 | -0.81 | -0.87 | -0.56 | -1.16 | -0.49 |
| Crops (compliant) | -0.11 | -0.10 | -0.11 | 0.23 | -0.34 | 0.35 | -0.37 | -0.10 | -0.11 |
| Livestock (non-compliant) | -0.58 | -1.07 | -0.26 | -0.47 | -0.63 | -0.50 | -0.61 | -0.85 | -0.44 |
| Livestock (compliant) | -0.11 | -0.12 | -0.10 | -0.03 | -0.17 | -0.16 | -0.08 | -0.12 | -0.10 |
| Forestry (non-compliant) | -1.59 | -2.01 | -1.33 | -1.52 | -1.62 | -1.61 | -1.51 | -2.27 | -1.21 |
| Forestry (compliant) | -0.34 | -0.34 | -0.35 | -0.18 | -0.46 | -0.23 | -0.44 | -0.35 | -0.33 |
| Fishing | 0.08 | 0.08 | 0.07 | 0.05 | 0.09 | 0.04 | 0.09 | 0.09 | 0.07 |
| Mining | -0.08 | -0.08 | -0.09 | -0.09 | -0.08 | -0.10 | -0.08 | -0.09 | -0.08 |
| Manufactures | -0.10 | -0.11 | -0.10 | -0.06 | -0.13 | -0.20 | -0.04 | -0.11 | -0.10 |
| Services | -0.12 | -0.12 | -0.12 | -0.11 | -0.13 | -0.12 | -0.12 | -0.13 | -0.11 |
| Total | -0.12 | -0.12 | -0.11 | -0.08 | -0.14 | -0.12 | -0.11 | -0.13 | -0.11 |
| **Labor Market** | | | | | | | | | |
| Real Wage | -0.36 | -0.38 | -0.35 | -0.27 | -0.43 | -0.34 | -0.37 | -0.40 | -0.34 |
| Unemployment Rate (pp.) | 0.00 | 0.00 | 0.00 | 0.00 | 0.00 | 0.00 | 0.00 | 0.00 | 0.00 |
| **Environmental** | | | | | | | | | |
| Deforestation (ha) | -2.45 | -35.27 | 15.03 | -9.27 | 0.97 | -2.59 | -2.38 | -3.28 | -2.00 |
| Deforestation Rate (pp.) | -0.03 | -0.42 | 0.18 | -0.11 | 0.01 | -0.03 | -0.03 | -0.04 | -0.03 |
| GHG emissions (mtCO2e) | -0.19 | -0.27 | -0.13 | -0.12 | -0.24 | -0.17 | -0.19 | -0.22 | -0.17 |
| **Land Use** | | | | | | | | | |
| Crops (non-compliant) | -0.71 | -2.82 | 0.67 | -0.23 | -0.95 | -0.90 | -0.55 | -1.20 | -0.48 |
| Crops (compliant) | -0.01 | 0.01 | -0.02 | 0.61 | -0.44 | 0.43 | -0.27 | 0.00 | -0.01 |
| Livestock (non-compliant) | -0.54 | -1.99 | 0.42 | -0.23 | -0.71 | -0.46 | -0.58 | -0.81 | -0.41 |
| Livestock (compliant) | -0.01 | -0.02 | -0.01 | 0.30 | -0.22 | -0.18 | 0.09 | -0.02 | -0.01 |
| Forestry (non-compliant) | -1.91 | -4.03 | -0.54 | -1.59 | -2.08 | -1.95 | -1.80 | -2.76 | -1.44 |
| Forestry (compliant) | -0.22 | -0.21 | -0.23 | 0.40 | -0.66 | -0.25 | -0.24 | -0.22 | -0.22 |

Own elaboration. S1= Deforested Land Supply Curve (LO); S2 = Deforested Land Supply Curve (UP); S3 = Land Wage Curve (LO); S4 = Land Wage Curve (UP); S5 = Transformation between domestic sales and exports (LO); S6 = Transformation between domestic sales and exports (UP); S7 = Transformation between exports to different destinations (LO); S8 = Transformation between exports to different destinations (UP). See descriptions in Table 3.

# 6. Discussion

Argentina is a middle-income country in Latin America that is significantly exposed to the EUDR, that will prohibit the export to the EU of certain products if they involve the use of deforested land. We estimate that the EUDR would cover around 6 billion US dollars in exported value, but only 2.84% is not compliant with the EUDR, being soy and cattle the most affected production chains.

We contributed to the literature on the implications of unilateral climate-related trade measures, such as the EUDR, in developing countries, with novel evidence on the macroeconomic, sectoral, and environmental impacts in Argentina. By using a computable general equilibrium model, we found that the potential macroeconomic impacts of the EUDR shock in Argentina are limited. As a consequence of the EU regulation, between 2025 and 2030, GDP would be reduced by an average of 0.14% with respect to the baseline scenario. However, some non-compliant sectors are highly impacted by the EUDR shock, such as soy, cattle, and timber production. The environmental impacts are higher: deforested hectares would be reduced by 2.45% and GHG emissions by 0.19%. These results depend on the modeling and data used. However, the main findings are robust, in particular, concerning their qualitative interpretation so that they account for the direction and range of magnitude of the effects, which are necessary as a basis for policy decisions aimed at mitigating the adverse effects.

Our study is not without limitations and potential future research agendas remain open. We model the EUDR shock with a focus on the export prohibition on non-compliant production. Notwithstanding, EUDR due diligence costs may still prevent compliant production from entering the EU market, so the total impacts could be higher. Delving into the uncertainties regarding compliance costs and studying what are the incentives and hurdles for compliance in each supply chain is an important area of future research (Cesar de Oliveira, 2024).

International prices are assumed to be exogenous and we do not consider changes as a consequence of the introduction of the EU regulation in the international market of the products covered. However, the literature suggests that world price effects would be expected in specific markets such as cocoa and coffee, where the EU accounts for a significant part of the global demand (Gilbert, 2024). Since those products are not significantly exported to the EU by Argentina, we consider that it should not affect our estimates.

As mentioned, we abstract from the part of the EU regulation that prohibits forest degradation, which is a more difficult process to measure and for which statistics are scarce (European Commission, 2021). We also abstract from considerations about the presence of segments of value chains with monopolistic competition that may reduce the effectiveness of this type of regulation (Dominguez-Iino, 2023). These are also potential future research agendas.

The EUDR is not the first nor the last unilateral climate-related trade regulation. Similar policies are under discussion in other countries of great relevance in the international market, such as the United States (Drost et al., 2022). Reducing deforestation is essential in the fight against climate change and Argentina should transform the challenges posed by these new regulations into opportunities to consolidate its position in international markets, comply with its environmental commitments, and improve practices in the agricultural sector, which is key to the climate transition and the country's development.

# Appendix A. International Trade Elasticities

**Table A1. International Trade Elasticities**

|  | Armington | CET |
|---|---|---|
| Animal and vegetable oils and fats (non-compliant) | 1.50 | 1.50 |
| Animal and vegetable oils and fats (compliant) | 1.50 | 1.50 |
| Live animals (non-compliant) | 2.00 | 2.00 |
| Live animals (compliant) | 2.00 | 2.00 |
| Sugar (non-compliant) | 2.00 | 2.00 |
| Sugar (compliant) | 2.00 | 2.00 |
| Wooden boxes and containers, cooperage products (non-compliant) | 1.50 | 1.50 |
| Wooden boxes and containers, cooperage products (compliant) | 1.50 | 1.50 |
| Meat and meat products (non-compliant) | 1.50 | 1.50 |
| Meat and meat products (compliant) | 1.50 | 1.50 |
| Cereals (non-compliant) | 2.00 | 2.00 |
| Cereals (compliant) | 2.00 | 2.00 |
| Cocoa, chocolate and confectionery (non-compliant) | 1.50 | 1.50 |
| Cocoa, chocolate and confectionery (compliant) | 1.50 | 1.50 |
| Leather and other leather articles (non-compliant) | 1.50 | 1.50 |
| Leather and other leather goods (compliant) | 1.50 | 1.50 |
| Waste or waste (non-compliant) | 0.90 | 0.90 |
| Waste or waste (compliant) | 0.90 | 0.90 |
| Newspapers, magazines and periodicals (non-compliant) | 1.50 | 1.50 |
| Newspapers, magazines and periodicals (compliant) | 1.50 | 1.50 |
| Prefabricated buildings (non-compliant) | 1.50 | 1.50 |
| Prefabricated buildings (compliant) | 1.50 | 1.50 |
| Crops of drinking plants and spices (non-compliant) | 2.00 | 2.00 |
| Crops of drinking plants and spices (compliant) | 2.00 | 2.00 |
| Live plants, flowers and seeds (non-compliant) | 2.00 | 2.00 |
| Live plants, flowers and seeds (compliant) | 2.00 | 2.00 |
| Fruits and nuts (non-compliant) | 2.00 | 2.00 |
| Fruits and nuts (compliant) | 2.00 | 2.00 |
| Residues from the extraction of vegetable fats; oilseed meals (non-compliant) | 1.50 | 1.50 |
| Residues from the extraction of vegetable fats; oilseed meals (compliant) | 1.50 | 1.50 |
| Legumes (non-compliant) | 2.00 | 2.00 |
| Legumes (compliant) | 2.00 | 2.00 |
| Printed products (except advertising) (non-compliant) | 1.50 | 1.50 |
| Printed products (except advertising) (compliant) | 1.50 | 1.50 |
| Record books, stationery, paper or cardboard (non-compliant) | 1.50 | 1.50 |
| Record books, stationery, paper or cardboard (compliant) | 1.50 | 1.50 |
| Wood (non-compliant) | 1.50 | 1.50 |
| Wood (compliant) | 1.50 | 1.50 |
| Works and carpentry pieces for construction (non-compliant) | 1.50 | 1.50 |
| Works and carpentry pieces for construction (compliant) | 1.50 | 1.50 |
| Furniture (non-compliant) | 1.50 | 1.50 |
| Furniture (compliant) | 1.50 | 1.50 |
| Oilseeds and fruits (non-compliant) | 2.00 | 2.00 |
| Oil seeds and fruits (compliant) | 2.00 | 2.00 |
| Other manufactures | 1.50 | 1.50 |
| Other agricultural products | 2.00 | 2.00 |
| Food products n.e.c. (non-compliant) | 1.50 | 1.50 |
| Food products n.e.c. (compliant) | 1.50 | 1.50 |
| Other animal products (non-compliant) | 2.00 | 2.00 |
| Other animal products (compliant) | 2.00 | 2.00 |
| Advertising material and other printed matter (non-compliant) | 1.50 | 1.50 |
| Advertising material and other printed matter (compliant) | 1.50 | 1.50 |
| Other products made of wood, cork and braidable materials (non-compliant) | 1.50 | 1.50 |
| Other products made of wood, cork and braidable materials (compliant) | 1.50 | 1.50 |
| Various basic chemicals (non-compliant) | 1.50 | 1.50 |
| Various basic chemicals (compliant) | 1.50 | 1.50 |
| Unprocessed vegetable materials n.e.c. (non-compliant) | 2.00 | 2.00 |
| Unprocessed vegetable materials n.e.c. (compliant) | 2.00 | 2.00 |
| Pulp, paper and cardboard (non-compliant) | 1.50 | 1.50 |
| Pulp, paper and cardboard (compliant) | 1.50 | 1.50 |
| Tires and other rubber products (non-compliant) | 1.50 | 1.50 |
| Tires and other rubber products (compliant) | 1.50 | 1.50 |
| Basic organic chemicals (non-compliant) | 1.50 | 1.50 |
| Basic organic chemicals (compliant) | 1.50 | 1.50 |
| Services related to agriculture, hunting, forestry and fishing (non-compliant) | 2.00 | 2.00 |
| Services related to agriculture, hunting, forestry and fishing (compliant) | 2.00 | 2.00 |
| Wood and other forestry products (non-compliant) | 2.00 | 2.00 |
| Wood and other forestry products (compliant) | 2.00 | 2.00 |
| Services | 0.90 | 0.90 |
| Raw tobacco (non-compliant) | 2.00 | 2.00 |
| Raw tobacco (compliant) | 2.00 | 2.00 |

Own elaboration.